\documentstyle[aps,prl,epsf,floats,amsmath]{revtex} 
\newcommand{\pd}{\partial}
\newcommand{\dd}{{\rm d}}
\newcommand{\e}{{\rm e}}
\newcommand{\bea}{\begin{eqnarray}}
\newcommand{\eea}{\end{eqnarray}}
\newcommand{\half}{\frac{1}{2}}

\draft

\begin{document}
\twocolumn[\hsize\textwidth\columnwidth\hsize\csname
@twocolumnfalse\endcsname

%\begin{flushright}
%UCLA/99/TEP/*
%\end{flushright}

\title{Solitosynthesis of $Q$-balls}

\author{Marieke Postma}
\address{Department of Physics and Astronomy, UCLA, Los Angeles, CA
90095-1547}
%\date{October 2001}

\maketitle  
             
\begin{abstract}
We study the formation of $Q$-balls in the early universe,
concentrating on potentials with a cubic or quartic attractive
interaction. Large $Q$-balls can form via solitosynthesis, a process
of gradual charge accretion, provided some primordial charge assymetry
and initial ``seed'' $Q$-balls exist.  We find that such seeds are
possible in theories in which the attractive interaction is of the
form $A H \psi^* \psi$, with a light ``Higgs'' mass.  Condensate
formation and fragmentation is only possible for masses $m_\psi$ in
the sub-eV range; these $Q$-balls may survive untill present.
\end{abstract}

\pacs{PACS numbers: 98.80.-k, 11.27.+d \hspace{1.0cm}
UCLA/01/TEP/31}

\vskip2.0pc]

\renewcommand{\thefootnote}{\fnsymbol{footnote}}

\setcounter{footnote}{0}

%%%%%%%%%%%%%%%%%%%%%%%%%%%%%%%%%%%%%%%%%%%%%%%%%%%%%%%%%%%%%%%%%%%%%%
%% Section I %%%%%%%%%%%%%%%%%%%%%%%%%%%%%%%%%%%%%%%%%%%%%%%%%%%%%%%%%
%%%%%%%%%%%%%%%%%%%%%%%%%%%%%%%%%%%%%%%%%%%%%%%%%%%%%%%%%%%%%%%%%%%%%%

\section{Introduction}

$Q$-balls are lumps of coherent scalar field that can be described
semi-classically as non-topological solitons.  They can arise in
scalar field theories with a conserved global $U(1)$ charge and some
kind of attractive interaction \cite{coleman}.  In a sector of fixed
charge, the $Q$-ball is the ground state of the theory.  $Q$-balls
generically occur in supersymmetric extensions of the standard model
\cite{mssm}. In these theories, baryon and lepton number play the role
of conserved charge.

$Q$-balls come in two types.  Type II $Q$-balls are associated with
the flat directions of the potential, which are a generic feature of
supersymmetric theories.  The VEV inside the $Q$-ball depends upon its
charge.  Formation of this type of $Q$-balls through fragmentation of
an Affleck-Dine (AD)-like condensate has been studied extensively in
the literature \cite{frag,em,ad}.  Type I $Q$-balls on the other hand
are characterized by a potential that is minimized at a finite VEV,
independent of the charge of the $Q$-ball.  We have analyzed under
which conditions this type of $Q$-ball can be formed in the early
universe. In this paper we present the results.

Large $Q$-balls can form via solitosynthesis, a process of gradual
charge accretion similar to nucleosynthesis, provided some primordial
charge assymetry exists~\cite{frieman,griest}.  The bottleneck for
this process to occur then is the presence of initial ``seed''
$Q$-balls.  Most potentials do not allow for small $Q$-balls which
makes solitosynthesis improbable.  The exceptions are theories in
which the attractive interaction is provided by a cubic term in the
Lagrangian of the form $ A H \psi^* \psi$, with a light ``Higgs''
mass.  Condensate formation does occur for light fields, for masses in
the range $m_\psi \lesssim {\rm eV}$.  If this leads to fragmentation,
the thus formed $Q$-balls can survive evaporation if their binding
energies are large.  $Q$-balls formed during a phase transition do not
survive evaporation, unless the phase transition takes place at
extremely low temperatures $T \lesssim 10^{-6} m_\psi$.

If $Q$-balls survive untill present they can be part of the dark
matter of the universe. Recently it was proposed that the dark matter
could be self-interacting; this would overcome various discrepancies
between observations and predictions based on collisionless dark
matter, such as WIMPs and axions~\cite{i_dm}.  Due to their extended
nature $Q$-balls have relatively large cross sections, and therefore
can naturally satisfy the required self-interactions~\cite{q_dm}.

Another cosmologically interesting feature of $Q$-balls is that
solitosynthesis in the false vacuum can result in a phase
transition~\cite{PT}.  Accretion of charge proceeds until a critical
charge is reached, at which point it becomes energetically favorable
for the $Q$-ball to expand, filling space with the true vacuum phase.

%%%%%%%%%%%%%%%%%%%%%%%%%%%%%%%%%%%%%%%%%%%%%%%%%%%%%%%%%%%%%%%%%%%%%%
%% Section II %%%%%%%%%%%%%%%%%%%%%%%%%%%%%%%%%%%%%%%%%%%%%%%%%%%%%%%%
%%%%%%%%%%%%%%%%%%%%%%%%%%%%%%%%%%%%%%%%%%%%%%%%%%%%%%%%%%%%%%%%%%%%%%

\section{$Q$-balls}

Consider a theory of an interacting scalar field $\phi$ that carries
unit charge under some conserved $U(1)$ charge.  The potential has a
global minimum $U(0)=0$ at $\phi = 0$.  We also require that the
potential admits $Q$-ball solutions, {\it i.e.},
\begin{equation}
	\mu_0 = \sqrt{ \frac{2U(\phi)}{\phi^2}} = {\rm min}, 
	\qquad {\rm for} \ \phi = \phi_0 \neq 0.
\label{mu_0}
\end{equation} 
The $Q$-balls solutions are of the form $\phi(\vec{x},t)={\rm
e}^{i\omega t}\bar{\phi}(\vec{x})$.  The charge and energy of such a
configuration is
\begin{equation}
	Q = \omega \int {\rm d}^3 x \ \bar{\phi}^2,
\label{Q}
\end{equation}
and
\begin{equation}
	E =  \int {\rm d}^3 x \ \left[ \half |\nabla \bar{\phi}|^2
	+ U_\omega( \bar{\phi}) \right] + \omega Q,
\end{equation}
with
\begin{equation}
	 U_\omega({\phi}) = U(\phi) - \half \omega^2 \phi^2.
\end{equation}
Minimizing the energy for a fixed $\omega$ is equivalent to finding a
3-dimensional bounce for tunneling in the potential $U_\omega$.  The
bounce solution exists for $\mu_0 < \omega < \sqrt{U''(0)}$ by virtue
of eq.~(\ref{mu_0}), and is spherically symmetric~\cite{coleman}.  The
equations of motion are:
\begin{equation}
	\frac{ \dd^2 \bar{\phi} }{ \dd r^2 } + \frac{2}{r} 
	\frac{ \dd \bar{\phi} } { \dd r }
	- \frac{ \pd U_\omega( \bar{\phi}) }{ \pd \bar{\phi} } = 0,
\label{bounce}
\end{equation}
with boundary conditions $\phi'(0)=0$ and $\phi(\infty)=0$.  

We will consider scalar potentials of the form
%
%\begin{subequations}
\begin{eqnarray}
	U_1(\phi) &&= \frac{1}{2} m_\phi^2 \phi^2 - A \phi^3 + \lambda
	\phi^4, 
\label{U1} \\
        U_2(\phi) &&= \frac{1}{2} m_\phi^2 \phi^2 - A \phi^4 + \lambda
	\phi^6;
\label{U2}
\end{eqnarray}
%\end{subequations}
%
both have $\phi_0 = A / 2 \lambda$ and $\mu_0^2 = m_\phi^2 - A^2 / 2
\lambda$.  $U_2$ is a typical potential that arises in effective field
theories.  $U_1$ is a non-polynomial potential, as the cube term is of
the form $(\phi^* \phi)^{3/2}$.  It is a typical potential in finite
temperature theories; this is however not interesting in the current
context since at high temperatures $Q$-balls evaporate quickly.  But
it can also arise as an effective field theory.  Consider for example
the potential
\bea
\label{U_1'}
	U'_1(\psi) =&& m_\psi^2 \psi^* \psi + m_H^2 H^* H -A' H \psi^*
	\psi +   {\rm h.c.}  \nonumber  \\ 
	&& +\frac{\lambda_1}{4} \psi^* \psi\, H^*\! H +
	 \hfill \frac{\lambda_2}{4} (\psi^* \psi)^2 
	+ \frac{\lambda_3}{4} (H^* H)^2,
\eea
where the ``Higgs'' field $H$ is uncharged under $U(1)$, whereas $\psi$
carries unit charge. Further, we take $A'$ real.  Now make the field
redefinitions
\begin{equation}
	{\rm Re} H =  \frac{1}{\sqrt{2}} \varphi \sin \theta, \qquad
	\psi = \frac{1}{\sqrt{2}} \varphi \e^{i \Omega} \cos \theta,
\end{equation}
then $U_1'$ becomes of the $U_1$ form, with $\phi$ some particular
direction in $(H,\psi)$-space.  We can also calculate $\mu_0^2 = 2
U_1'/(\varphi^2 \cos^2 \theta) $ in terms of the $U_1'$
parameters. Taking $m_H =0$ and all quartic couplings equal $\lambda_1
= \lambda_2 = \lambda_3 = \lambda'$ to simplify the algebra, this
yields
\begin{equation}
	\mu_0^2 = m_\psi^2 - \frac{ {A'}^2}{3 \lambda'}  
\label{mu}
\end{equation}
at $\theta_0 = \pi/4$ and $\varphi_0 = 4 A'/ 3 \sqrt{2} \lambda'$. 

Potentials of the form $U'_1$ are present in the scalar sector of the
MSSM, where the Higgs field couples to sparticle fields through a
cubic interaction \cite{mssm}.  The sparticles carry a conserved
$U(1)$ charge in the form of baryon or lepton number.  However, the
sparticles and also possibly formed $Q$-balls are unstable, as they
can decay into light fermions \cite{evap_ferm}.  Stable $Q$-balls can
be obtained in a model where the standard model (SM) Higgs field is
coupled to a charged SM singlet~\cite{demir}.  The SM singlet $\psi$
is charged under a hidden sector $U(1)_\psi$ global symmetry, under
which none of the SM particles are charged.  The $Q$-balls in this
model interact with the SM particles only weakly, through the Higgs
boson.  Another possibility is that both the $H$ and $\psi$ field are
hidden sector fields, interacting only gravitationally or through some
other surpressed interaction with the visible sector.  Hidden sectors
appear in a variety of models, such as technicolor, mirror symmetry,
hidden sector SUSY breaking, and brane world models. They also arise
naturally from string theory; in heterotic $E_8 \times E_8$ string
theory compactified on a Calabi-Yau manifold, one of the $E_8$'s
contains the SM whereas the other is some hidden sector.

We will assume an initial charge asymmetry, {\it i.e.}, an excess of
particles over anti-particles.  This asymmetry may be created through
a mechanism similar to those invoked to explain the baryon asymmetry
in the universe, such as the Affleck-Dine mechanism~\cite{affleck}.

%%%%%%%%%%%%%% subsection %%%%%%%%%%%%%%%%%%%%%%%%%%%%%%%%%%%%%%

\subsection{Large $Q$-balls --- thin wall approximation}

For large $Q$ the $Q$-ball solution can be analyzed using a thin wall
approximation, which consists of neglecting the effect of the surface
compared to the bulk.  The $Q$-ball may be approximated by a sphere of
radius $R_Q$ with $\phi = \phi_0$ inside and zero field value outside.
The mass and radius of the solition are
\begin{equation}
	M_Q = \mu Q,
\end{equation}
and
\begin{equation}
	R_Q = \frac{\beta_Q}{m_\phi} Q^{1/3}, 
	\qquad \beta_Q = \left( \frac{3 m_\phi^3}
	{4 \pi \omega \phi_0^2} \right)^{1/3},
\label{R_large}
\end{equation}
with $\mu,\omega \to \mu_0$ for $Q \to \infty$.  The soliton is large
and its cross section is given by the geometrical area
\begin{equation}
	\sigma_Q = \pi R_Q^2.
\end{equation}

%%%%%%%%%%%%%% subsection %%%%%%%%%%%%%%%%%%%%%%%%%%%%%%%%%%%%%%

\subsection{Small $Q$-balls}
\label{sec_small}

The limit of small charge corresponds to $\omega \to m_\phi$. In this
limit the solution of the bounce equation (\ref{bounce}) is of the
form \cite{fin}
\begin{equation}
	\bar{\phi} \sim (m_\phi^2-\omega^2)^{2-a} \,
	\e^{-\sqrt{m_\phi^2-\omega^2} r},
\end{equation}
with $a$ the power of the attractive term in the potential.  This
solution has the right behavior for $r \to \infty$ where $\phi \to 0$
and the quadratic term in the effective potential dominates, and for
$\omega \to m_\phi$ where the zero of $U_\omega$ is near the origin.
Using the solution to compute the conserved charge (\ref{Q}), and
taking the limit $\omega \to m_\phi$, one finds a finite, non-zero
value only for $4 + 2D - aD > 0$, with $D$ the number of spatial
dimensions.  In $D=3$ dimensions, $U_1$ admits small $Q$-balls but
$U_2$ does not.  Therefore, we will only consider $U_1$ in the
remaining of this section.

In the limit of large $\omega$, or equivalently very non-degenerate
minima, one can neglect the quartic terms in $U_\omega(\phi)$.  This
is the thick wall approximation \cite{thick}.  The approximation is
valid for $Q$-balls with charge $Q$ that satisfies, for potential
$U_1$:
\begin{equation}
	\left \{
	\begin{array}{ll}
		&Q \ll 14.6 m_\phi / \sqrt{\lambda}  A, \\
		&Q < 7.3 m_\phi^2 / A^2.
	\end{array}
	\right.
\label{cond_thick}
\end{equation}
If above conditions are met one can define an expansion parameter
\begin{equation}
	\epsilon \equiv Q \frac{A^2}{3 S_\psi m_\phi^2} < \frac{1}{2},
\label{epsilon}
\end{equation}
with $S_\psi \approx 4.85$.  The mass of the soliton is
\begin{equation} 
	M_Q = Q m_\phi \left[ 1 - \frac{1}{6} \epsilon^2 + {\cal O}
	(\epsilon^4) \right].
\label{mass_small}
\end{equation}
The radius of the $Q$-ball can be parameterized 
\begin{equation}
	R_Q = \frac{\beta_Q}{m_\phi} Q^{1/3},
\label{R_small}
\end{equation}
with $\beta_Q \sim {\mathcal O}(1)$.

The $Q$-balls described above are classically stable for arbitrary
small charge $Q$.  However, one expects quantum fluctuations to become
important in this regime. Indeed, numerical calculations indicate that
this is the case, and only configurations with $Q \gtrsim 7$ are
quantum mechanically stable \cite{graham}.  All these calculations are
based on the potential $U_1$.  This potential is an effective
potential which is well suited for a semi-classical description of
large $Q$-balls.  But for small $Q$-balls the degrees of freedom of
the underlying potential $U_1'$ should be taken into account.  In this
regime a treatment in terms of quantum bound states is more
appropiate.  Solving the bound state problem in full generality is not
an easy task.  However in the limit that all quartic interactions can
be neglected, the theory becomes identical to the Wick-Cutkosky
model. This model can be solved analytically for the case of a
massless exchange particle, {\it i.e.}, $m_H=0$.  The various
approaches used in the literature, {\it e.g.} ladder approximation,
Feshback-Villars formulation, variational-perturbative calculations
\cite{wc,ladder,crit}, all lead to the same result that the bound
state spectrum is discrete with the $n^{\rm th}$ state having an
energy (to lowest order in $\alpha$):
\begin{equation}
	E_n = 2 m_\psi (1 - \frac{\alpha^2}{8n^2}),
	\qquad \alpha = \frac{1}{16\pi} \frac{{A'}^2}{m_\psi^2}.
\label{alpha}
\end{equation}

The above result for the binding energy is derived in the limit of a
massless boson exchange. No analytic results are known for massive
scalar exchange.  However, numerical studies show that bound states
still form, provided that $\alpha$ is larger than some critical value.
We estimate, based on the results in~\cite{crit}, that bound states
exist for
\begin{equation}
	\alpha > \alpha_{\rm min} \approx 2 \frac{m_H}{m_\psi} +
	{\mathcal O} \left( \frac{m_H}{m_\psi} \right)^2.
\end{equation}
That is, the Higgs mass needs to be sufficiently small $m_H \lesssim
10^{-2} (A'/m_\psi)^2 m_\psi$. The energy of the bound state is of the
same parametric form as for the massless case.

The other assumption that went into the derivation of eq.
(\ref{alpha}) is the absence of quartic couplings.  We expect this to
be a good approximation in the regime where quartic interactions are
negligible small.  The cross section for $\psi \psi$-scattering is
$\sigma_{\psi \psi \to \psi \psi} = {S |{\mathcal M}|^2}/ {16 \pi
E_{\rm cm}^2}$.  For scattering through Higgs exchange, governed by
the cubic interaction, this gives at tree level
\begin{equation} 
	\sigma^{\rm cubic} \approx \frac{1}{128 \pi
	m_\psi^2} \frac{{A'}^4}{E^4} \; \stackrel{T \lesssim
	A'} {\longrightarrow} \; \sim \pi R_\psi^2.
\label{sigma_psi}
\end{equation}
Here $E = {\rm max}\{T, m_H\}$.  At low temperatures $T \lesssim A'$,
which are the temperatures of interest, the cross section quickly
approaches the unitarity bound and higher order diagrams cannot be
neglected.  In this regime we will approximate the cross section by
$\sigma \sim \pi R_\psi^2$ with $R_\psi = 2 \pi/m_\psi$ the Compton
wavelength.  Scattering through the quartic point interaction has an
amplitude $|{\mathcal M}| = \lambda'$. And thus the requirement that
the repulsive quartic interactions are negligible small $\sigma^{\rm
quartic} \ll \sigma^{\rm cubic}$, are fullfilled for all quartic
couplings ${\lambda'} \lesssim 1$.  It may be that also for
non-perturbative values of the quartic couplings bound states persist;
but this certainly cannot be \nobreak{analysed} perturbatively. As it
seems unnatural to have quartic couplings larger than one, we will
ignore this possibility.

On the other hand, the quartic couplings cannot be arbitrary small or
else no $Q$-ball solution exists: for the case of zero Higgs mass and
all quartic couplings equal, $\mu_0^2 > 0$ translates into $\lambda' >
{A'}^2/ 3 m_\psi^2$, as follows from
eq. (\ref{mu}).~\footnote[2]{Condition $\mu_0^2 > 0$ corresponds to the
requirement that $\phi=0$ is the global minimum of the potential.
$Q$-ball solutions do exist for $\phi=0$ a local minimum.  In the
potentials $U_1$ and $U_2$ this possibility is not realized, since at
low temperatures the field will end up in the true vacuum.  ($U_1$: at
the temperature $T$ that the minimum at $\phi \neq 0$ becomes global
the energy barrier is $\sim 10^{-2} \lambda T^4$.  $\; U_2$: at high
temperature $m^2(T) < 0$.)}  Non-zero, but small Higgs mass $m_H <
10^{-2} m_\psi$ does not alter this result noticeably.  The quartic
couplings do not have to be all equal, but at least one of them has to
be ${\mathcal O}({A'}/ m_\psi)^2$. For \nobreak{$A' = m_\psi$},
$Q$-ball solutions exist for example for
$(\lambda_1,\lambda_2,\lambda_3) = (0.4, \,0.4, \,0.4), \; (1, \,0.01,
\,0.01)$ and $(0.05, \,0.8, \,0.3)$.

Both the quantum bound states discussed above and $Q$-balls describe
the same objects --- stable bound states with a fixed global charge
--- but in a different language. In both descriptions the existence
and stability of the bound state relies on the trilinear coupling and
the conserved global charge (that is, conserved particle number).  For
large bound states quantum fluctuations can be neglected, and a
semi-classical description as a $Q$-ball becomes a good approximation.
The trilinear coupling makes it possible for the energy of a bound
state with a fixed charge to be less than a collection of free
particles with the same charge. In the limit of small particle number
(global charge), it becomes necessary to treat the full quantum
problem because the semiclassical approximation breaks down.  The
trilinear term can be viewed as an attractive interaction between the
$\phi$-particles, which makes it possible for bound states to form.
The lowest level bound state is the stable ground state, as charge
conservation forbids it to lower its energy through annihilation of
$\phi$ particles.

It is tempting to compare the ground state result ($n = 1$) of
eq.~(\ref{alpha}) with the $q = 2$ result obtained in the thick wall
approximation (\ref{mass_small}): both mass formulas give the same
parametric dependence.  However, in the overlapping regime both
approximations are taken beyond their domain of validity: for equal
masses $m_H = m_\psi$ bound states can only form for large $\alpha$,
and for $q=2$ a semi-classical treatment breaks down.  Of course both
approximations are similar in that they neglect the quartic
interactions.

In conclusion, the potential $U_1'$ admits stable, two-particle bound
states at low temperatures (below the binding energy), provided the
Higgs mass is sufficiently light, and the quartic repulsive
interactions small.  We repeat that our assumption here is that
non-zero quartic couplings do not destabilize the bound state provided
$\sigma^{\rm quartic}_{(\phi \phi \to \phi \phi)} \ll \sigma^{\rm
cubic}_{(\phi \phi \to \phi \phi)}$; this should be checked by an
explicit calculation. For the potential to have a global minimum at
$\phi =0$, or equivalently for $Q$-ball solutions to exist into which
the bound states can grow, the couplings cannot be too small.
\begin{eqnarray}
	\left\{ 
	\begin{array}{lll}
		& {\lambda'} \lesssim 1
		& \quad {\rm repulsive} \: {\rm forces} 
		\:  {\rm small} \\
		& {m_H}  \lesssim 10^{-2}  \left( 
		\frac{{A'}^2}{m_\psi^2} \right) {m_\psi}
		& \quad {\rm small} \: {\rm Higgs} \: {\rm mass}\\
		& \lambda' > \frac{{A'}^2} {3 m_\psi^2} 
		& \quad Q{\rm -balls} \: {\rm exist}
	\end{array}
	\right.
\end{eqnarray}
A possible set of parameters is $\lambda' \sim 0.5$, $A' \sim m_\psi$
and $m_H \lesssim 10^{-2} m_\psi$.  The binding energy for the bound
state is then $B_2 = \alpha^2/8 \sim 5 \times 10^{-5} m_\psi$, and
$\mu_0 \sim 0.6 m_\psi$.

We will further assume that similary bound states of more than two
particles can exist, and that they have energies
\begin{equation}
	M_Q = Q m_\psi (1 -  f_{_Q}\frac{\alpha^2}{8}),
	\qquad \alpha = \frac{1}{16\pi} \frac{{A'}^2}{m_\psi^2}.
\label{M_bnd}
\end{equation}
with $f_{_Q}$ some unknown factor depending on the charge $Q$, the
mass of the exchange particle and the strength of the quartic
interactions.

The binding energy of a $Q$-ball is $B_Q = Q m_\psi -
M_Q$. Two-particle bound states are only stable at temperatures below
the binding energy $T < B_2 \sim \alpha^2/8$. From then on they can
grow by capturing charged particles. A non-relativistic particle with
kinetic energy $E_k \sim T$ has energy $T + B_Q$ inside the
$Q$-ball/bound state.  For it to be captured it has to lose an amount
larger than $T$ in the collision.  Assuming isotropy, on average a
particle will lose half of its energy.  Therefore, for temperatures $T
< B_Q$ a considerable amount of the particles scattering with the
$Q$-ball will be captured. We approximate the absorption cross section
$\sigma_{\rm abs}(Q)$ for a $Q$-ball with charge $Q$ by the scattering
cross section: $\sigma_{\rm abs}(Q) \sim \pi R_Q^{2}$.

%%%%%%%%%%%%%%%%%%%%%%%%%%%%%%%%%%%%%%%%%%%%%%%%%%%%%%%%%%%%%%%%%%%%%%
%% Section III %%%%%%%%%%%%%%%%%%%%%%%%%%%%%%%%%%%%%%%%%%%%%%%%%%%%%%%%
%%%%%%%%%%%%%%%%%%%%%%%%%%%%%%%%%%%%%%%%%%%%%%%%%%%%%%%%%%%%%%%%%%%%%%

\section{Solitosynthesis}

In thermal equilibrium, the production of large $Q$-balls through
gradual charge acretion is very efficient.  This process is dubbed
solitosynthesis for its similarity with nucleosynthesis.  It requires
an initial charge asymmetry not unlike the baryon asymmetry of the
universe.  Freeze out of any of the reactions involved will put a halt
to solitosynthesis.

In this section we will describe the thermodynamics of $Q$-balls in
terms of a gas of non-relativistic $\psi$ particles in thermal
equilibrium.  The $\psi$ particles can bind together through the
exchange of a light scalar particle, as given by the cubic interaction
in $U'_1$.  For large $Q$-balls a semi-classical description in terms
of $U_1$ suffices, and $\psi$ can be replaced by $\phi$ in all the
formulas.

%%%%%%%%%%%%%%%%%% subsection %%%%%%%%%%%%%%%%%%%%%%%%%%%%%%%%%%%%

\subsection{$Q$-balls in thermal equilibrium}

At non-relativistic temperatures $T < m_\psi$, the number densities of
a distribution of $Q$-balls and free $\psi$ particles in thermal
equilibrium are governed by the Boltzmann distribution:
\begin{equation}	
	n_Q(T) = g_Q \left( \frac{M_Q T}{2 \pi} \right)^{3/2} 
	\e^{   {(\mu_Q - M_Q)}/{T} }, 
\end{equation}
and 
\begin{equation}
	n_\psi(T) = g_\psi \left( \frac{m_\psi T}{2 \pi} \right)^{3/2} 
	\e^{(\mu_\psi - m_\psi)/{T}}.  
\label{n_phi}
\end{equation} 
Here $g_Q$ is the internal partition function of the $Q$-ball, and
$g_\psi = 2$, the number of degrees of freedom of a complex field.
Solitosynthesis is only possible if capture rates are large compared
to the expansion rate of the universe, otherwise the densities are
frozen.  If so, the gas of $\psi$ particles and $Q$-balls is in
chemical equilibrium, and the accretion and absorption reactions
\begin{equation}
	(Q) + \psi \longleftrightarrow (Q+1)
\label{reaction}
\end{equation}
enforce a relation between the various chemical potentials: $ \mu_Q =
Q \mu_\psi$.  This allows one to express the $Q$-ball number density
in terms of the $\psi$-number density
\begin{equation}	
	n_Q(T) = \frac{g_Q}{g_\psi^Q} n_\psi^Q
	\left( \frac{M_Q}{m_\psi} \right)^{3/2}
	\left( \frac{2 \pi}{m_\psi T} \right)^{3(Q-1)/2} 
	\e^{{B_Q}/{T} }, 
\label{nq}
\end{equation}
with $B_Q = Q m_\psi - M_Q > 0$ the binding energy of a $Q$-ball.
Similar equations can be written down for the number densities of
anti-$\psi$'s and anti-$Q$-balls.

We will assume a primordial asymmetry of $\psi$'s over $\psi^*$'s,
$\eta \equiv (n_\psi - n_{\psi^*}) / n_\gamma$, where $n_\gamma = 2.4
T_\gamma^3 / \pi^2$ is the photon number density. Initially one has
both $\psi$ and $\psi^*$ particles.  For large asymmetry the
anti-particle density can be neglected.  Also, if the Higgs mass is
light then pair annihilation occurs, and at non-relativistic
temperature anti-particles deplete rapidly.  The annihilation reaction
enforces $\mu_\psi = -\mu_{\psi^\ast}$, which in the non-relativistic
limit leads to
\begin{equation}
	n_{\psi^\ast} = n_\psi \e^{-2\mu_\psi/T}.
\end{equation}
For temperature $T \lesssim m_\psi$ the chemical potential $\mu \sim
m_\psi$; otherwise the Boltzmann suppression $\exp[(\mu - m) / T]$ is
tremendous and the charge conservation equation
\begin{equation}
	\eta n_\gamma = n_\psi - n_{\psi^\ast} + \Sigma Q n_Q +
	\Sigma Q^\ast n_{Q^\ast}.
\label{charge}
\end{equation}
can never be satisfied.  Annihilation is efficient untill the density
of anti-particles is negligible small.  The number density of stable
$\psi$-particles is then
\begin{equation} 
	n_\psi \approx \eta n_\gamma, 
	\quad \eta  = 2.5 \times 10^{-8} \Omega_\psi h^2 
	\frac{{\rm GeV}}{m_\psi}.
\label{eta}
\end{equation}
The local density can be higher if clustering occurs.  This is not to
be expected until the matter dominated era, $T < T_{\rm eq} \approx
5.5 (\Omega_0 h^2)^{-1} \zeta^{-1} {\rm eV}$.

The photon temperature may in general be different from the
temperature of the $Q$-ball system.  Particle species that decouple
from the heat bath when they are highly relativistic maintain an
equilibrium distribution with temperature $T \propto R^{-1}$.  The
photon temperature red shifts as $T_\gamma \propto g_{* s}^{-1/3}
R^{-1}$, and thus the difference in temperatures is given by a factor
$\zeta \equiv (g_{* s}(T_D) - g_{* s}(T))^{1/3}$, with $T_D$ the
temperature at which the $Q$-ball system decouples. When the
$\psi$-particles only interact gravitationally $\zeta \sim 10$,
whereas it can be much lower for more general interactions $\psi
\psi^\ast \longleftrightarrow X$, where $X$ are light particles that
do not carry the same $U_Q(1)$ charge as the $\psi$-particles.  We
parameterize $T_\gamma = \zeta T$, with $\zeta \sim 1 - 10$.

\begin{figure}[t]
\centering
\hspace*{-5.5mm}
\leavevmode\epsfysize=6cm \epsfbox{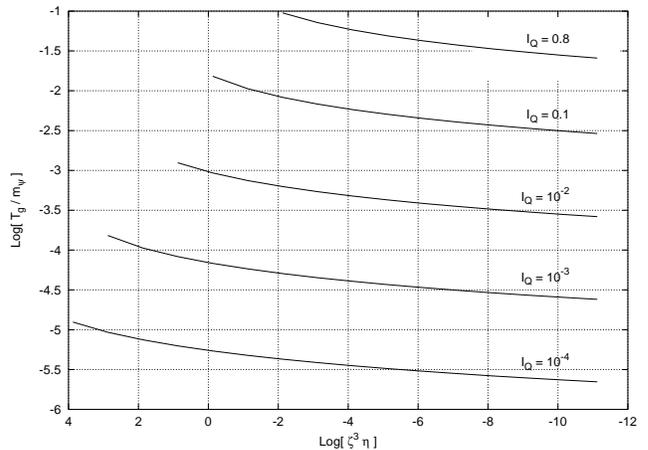}\\[3mm]
\caption[fig.2] {Accretion time $T_g$ plotted as a function of
$c\zeta^3 \eta$ for various values of $I_Q/m_\psi$. For large values of
$c\zeta^3 \eta$ --- to the left of where the lines stop --- accretion
dominates over evaporation at all (non-relativistic) temperatures}
\end{figure}

The $Q$-ball densities can start growing when the exponent in
equation~(\ref{nq}) dominates over the potentially small factor in
front.  Since $B_Q$ grows with $Q$, formation of large $Q$-balls is
favored. The evolution of a single $Q$-ball is given by the absorption
and evaporation rates of $\psi$ particles by a $Q$-ball of charge $Q$
\cite{frieman}. These can be found using detailed balance
arguments. In chemical equilibrium we have for the process in
eq.~(\ref{reaction})
\begin{equation}
	n_\psi \, v_\psi \, \sigma_{\rm abs} (Q) 
	= n_{Q+1} \, r_{\rm evap}(Q+1),
\label{balance}
\end{equation}
and
\begin{eqnarray}	
	\frac{{\rm d} Q}{{\rm d} t} 
	&=& r_{\rm abs}(Q) - r_{\rm evap}(Q) \nonumber \\
	&=& n_\psi v_\psi [ \sigma_{\rm abs}(Q) 
	- \frac{n_{Q-1}}{n_Q} \sigma_{\rm abs}(Q-1)].
\end{eqnarray} 
The $Q$-ball starts growing when $r_{\rm abs}(Q) > r_{\rm evap}(Q)$.
This happens for $T \lesssim T_g$ with
\begin{equation}
	\frac{T_g}{m_\psi} = \frac{I_Q / m_\psi}
	{- \frac{3}{2} \ln \left( \frac{T_g}{m} \right) 
	- \ln \left (c \zeta^3 \eta
	\right)},
\label{T_g}
\end{equation} 
and
\begin{equation} 
	I_Q = m_\psi + M_{Q-1} - M_{Q}.  
\end{equation} 
Here $c= (Q/Q-1)^{(13/6)} g_Q/g_{Q-1}$, which goes to one for large
$Q$, and $c \sim 10$ in the limit $Q\to 2$.  $I_Q$ is the binding
energy with which a single $\psi$-particle is bound to the
$Q$-ball. For very small $Q$-balls $I_Q / m_\psi = f_{_Q} \alpha^2/8$.
Accretion of the smallest $(Q_{\rm min})$-ball starts when $T \sim
I_{Q_{\rm min}} < 10^{-6} m_\psi$ for $A' \lesssim 1$.  For large
$Q$-balls $I_Q = m_\psi - \mu_0$.  Figure~1 shows $T_g$ as a function
of $c \zeta^3 \eta$ for various values of $I_Q$.  For large $c \zeta^3
\eta$ equation ({\ref{T_g}) has no solution; here absorbtion dominates
at non-relativistic temperatures.

For large $H$ mass and early freeze out the charge asymmetry may be
small, as then both $\psi$ and $\psi^\ast$-densities are large at
freeze out while annihilation is negligible. Both particle and
anti-particle number are conserved, and one can in principal have
growing $Q$- and anti-$Q$-balls at the same time.  In this case
however, the formation of small seed $Q$-balls, which are necessary to
start the fusion process, appears to be a major obstacle.

%%%%%%%%%%%%%%%%% subsection %%%%%%%%%%%%%%%%%%%%%%%%%%%%%%%%%%%%%%

\subsection{Freeze out}

For solitosynthesis to work $T_g$ must be higher than $T_{\rm FO}$,
the temperature at which the absorption reactions eq.~(\ref{reaction})
freeze out.  This occurs when the reaction rate for accretion becomes
smaller than the expansion rate of the universe:
\begin{equation}
	\Gamma[(Q) + \psi \to (Q+1)] \lesssim H.
\label{freeze}
\end{equation}
The Hubble constant during the radiation dominated era is $H = 1.7
g_{* s}^{1/2} T_\gamma^2 / M_{\rm pl}$; it is $H = 1.7
g_{* s}^{1/2} T_\gamma^{3/2} T_{\rm eq}^{1/2} / M_{\rm pl}$ in
the matter dominated era. The effective degrees of freedom are $g_{*
s} \lesssim 10$ for $T_\gamma \lesssim 10 {\rm MeV}$ and $g_{* s} \sim
100$ for $0.1 {\rm GeV} \lesssim T_\gamma \lesssim 10^3 {\rm GeV}$.
The accretion rate is $\Gamma = n_\psi v_\psi \sigma_{\rm abs}(Q)$.
Neglecting the charge density subsiding in $Q$-balls, then for stable
$\psi$ particles $n_\psi$ is given by eq. (\ref{eta}). At
non-relativistic temperatures $v_\psi =(T/2\pi m_\psi)^{1/2}$.

We are interested in temperatures $T<T_g$; then the cross section is
$\sigma_{\rm abs} \sim \pi R_Q^2$ and freeze out occurs for
temperatures, for $T_{\rm FO} > T_{\rm eq} \approx 5.5 (\Omega_0
h^2)^{-1} \zeta^{-1} {\rm eV}$:
\begin{equation}
	\frac{T_{\rm FO}}{m_\psi}  \lesssim 
	\left[ \frac{10^{-9}}{ \zeta \beta_Q  Q^{2/3}}  
	\left(\frac{m_\psi}{{\rm GeV}}\right)^2
	\left(\frac{0.3}{\Omega_\psi h^2}\right)
	\left(\frac{g_*^{1/2}}{10}\right)
	\right ]^{2 / 3}. 
\end{equation}
And for $T_{\rm FO} < T_{\rm eq}$:
\begin{equation}
	\frac{T_{\rm FO}}{m_\psi}  \lesssim
	\left[ \frac{10^{-13}}{ \zeta \beta_Q  Q^{2 / 3}}  
	\left(\frac{m_\psi}{{\rm GeV}}\right)^{3/ 2}
	\left(\frac{0.3}{\Omega_\psi h^2}\right)^{1 / 2}
	\left(\frac{g_*^{1/2}}{10}\right)
	\right ]^{1 / 2}.
\end{equation}
For $A' \sim m_\psi$, $\zeta \sim 10$ and $m_\psi \lesssim {\rm GeV}$
freeze out of the accretion reactions for the smallest $Q$-balls ($Q
=2$) occurs after the accretion phase, $T_{\rm FO} \lesssim T_g$.  In
this case solitosynthes can start at photon temperature $T_\gamma =
\zeta T_g \sim 10^{-5} m_\psi \lesssim 10^4 {\rm eV}$.  Note that $I_Q
\propto (A'/m_\psi)^4$ decreases rapidly for smaller cubic couplings
and $T_{\rm FO} < T_g$ can only be satisfied for increasingly low
$\psi$-mass. In the matter dominated era the reaction rate, and thus
the freeze out temperature, can be increased through clustering.  For
an overdensity of $\sim 10^5$ in galaxies, we find that for $A' \sim
0.1 m_\psi$ solitosynthesis occurs for $m_\psi \lesssim {\rm GeV}$,
starting at temperatures $T_\gamma \lesssim {\rm eV}$. For smaller
values $A' \lesssim 0.1$ or $m_\psi \lesssim {\rm MeV}$,
solitosynthesis occurs in the future, at temperatures smaller thant
the present day temperature $T_\gamma < T_0 = 2.35 \times 10^{-4} \,
{\rm eV}$.

With the accumulation of charge in $Q$-balls the number density of
$\psi$ particle decreases and the system freezes out.  Since the
accretion is such an explosive process, this will generelly not happen
until almost all charge resides in $Q$-balls. More quantitatively,
when the $\psi$-density decreases to 10\% of its original value,
$T_{\rm FO}$ decreases only by a factor $10^{-2/3}$.  The back
reaction is only important when $T_g \approx T_{\rm FO}$, and it shuts
off the growth of $Q$-balls immediately; in all other cases most
charge will end up in $Q$-balls.

The accretion rate of a {\rm single} $Q$-ball is limited by the
diffusion rate.  However, diffusion of charge is only important when
$l \lesssim R_Q$, with $l \sim \Gamma_{\psi \psi}^{-1}$ the mean free
path.  The radius $R_Q$ of a $Q$-ball becomes equal to the mean free
path for a large charge:
\begin{equation}
	Q_{\rm diff}(T) \sim 10^{87}
 	\left[ \left( \frac{m_\psi/T}{10^7} \right)^{7/2}
        \left( \frac{10}{\zeta} \right)^3 \frac{1}{\beta}
	\frac{0.3}{\Omega h^2} \frac{m_\psi}{{\rm GeV}} \right]^3
\end{equation}
For $Q> Q_{\rm diff}$ diffusion is important.  The total amount of
charge inside a Hubble volume is $Q_{\rm total} = n_\psi H^{-3}$
\begin{equation}
	Q_{\rm total} (T)
	\sim 10^{63} \left(\frac{m_\psi/T}{10^7}\frac{10^2}{\zeta g_*^{1/2}} \right)^3
	\left( \frac{{\rm GeV}}{m_\psi}\right)^4
	\left( \frac{\Omega_\psi h^2}{0.3} \right)
\label{Q_total}
\end{equation}
For small masses
$Q_{\rm diff}$ may be lower than the total charge inside a Hubble
volume; in this case $Q_{\rm diff}$ will be an upper limit on the
charge of the $Q$-balls formed during solitosynthesis.

%%%%%%%%%%%%%%%%%%%%%%%%%%%%%%%%%%%%%%%%%%%%%%%%%%%%%%%%%%%%%%%%%%%%%%%
%% Section IV %%%%%%%%%%%%%%%%%%%%%%%%%%%%%%%%%%%%%%%%%%%%%%%%%%%%%%%%
%%%%%%%%%%%%%%%%%%%%%%%%%%%%%%%%%%%%%%%%%%%%%%%%%%%%%%%%%%%%%%%%%%%%%%

\section{Seeds}

Solythosynthesis is a very effecient way to form large $Q$-balls,
provided there are some initial seed $Q$-balls at temperatures above
freeze out.  These seeds may be remnants of an earlier epoch, formed
during a phase transition or via the decay of a Bose-Einstein
condensate. Another option is that small stable $Q$-balls can form in
the gas of $\phi$-particles.

%%%%%%%%%%%%%%%%% subsection %%%%%%%%%%%%%%%%%%%%%%%%%%%%%%%%%%%%

\subsection{Formation of small $Q$-balls}

As discussed in section \ref{sec_small}, small $Q$-ball solutions are
only stable for potentials with a cubic interaction.  Two-particle
bound states can form through scalar exchange, provided the mass of
the exchange boson is sufficiently small and the quartic interactions
can be neglected.  In this case seed $Q$-balls can be formed copiously
and solitosyntesis can start.  If the mass of the scalar mass is of
the same order as the mass of the charged particles two-particle bound
states do not form, but it may still be that small $Q$-balls with
charge $Q_{\rm min} > 2$ are stable.  Numerical calculations indicate
that in the thick wall approximation (which has $m_H =m_\psi$)
$Q$-balls are quantum mechanically stable for $Q_{\rm min} \gtrsim 7$
\cite{graham}.

For $Q > 2$, $Q$-ball formation is surpressed compared to the
two-particle bound state, by the requirement that $Q$ charges should
be similtaneously in a volume of radius $\sim R_q$.  Define $P(q)$ to
be the probability to find a charge $Q$ in the volume of a $Q$-ball,
$V_q \approx R_q^3$.  The mean charge in $V_q$ is $\bar{q} = n_\psi
V_q$, whereas the variance is $\sigma^2 = \langle (\Delta q)^2 \rangle
= T ( \partial \bar{q} / \partial \mu)_{T,V} = \bar{q}$. Since
\begin{equation}
	\bar{q} \approx 1.0 \, \zeta^3 \eta q 
	\left( \frac{T}{m_\psi} \right) \ll 1
\end{equation}
a discrete distribution is needed, the Poisson distribution:
\begin{equation}
	P(q) = \frac{{\rm e}^{-\bar{q}} \bar{q}^{q}}{q!} 
	\approx \frac{(n_\psi V_q)^{q}}{q!}.
\end{equation}
The density of lumps with charge $Q$ in a volume $V_q$ is $n_{q} =
P(q) / V_q$.  The reaction rate for the bound state is $\Gamma^{\rm
bnd}_q \sim \sigma^{\rm bnd}_q n_{q}$ so that the chance that in a
Hubble time a ``$Q$-lump'' forms a bound state is $\sim n_{q}
\sigma^{\rm bnd}_q H^{-1}$.  Multiplying this with the total number of
$Q$-lumps in a Hubble volume gives the number of $Q$-ball seeds $N_q
\sim n_{q}^2 \sigma^{\rm bnd}_q H^{-4}$.  Taking $R_Q \sim 1/m_\psi$
this yields
\begin{equation}
	N_q \sim \eta^q (\sigma_{\rm bnd} m_\psi^2) 
	\left(\frac{T}{m_\psi}\right)^{6Q-8} 
	\left( \frac{M_{\rm pl}}{m_\psi} \right)^4.
\end{equation}
Assuming $\sigma^{\rm bnd}_q \lesssim \sigma_{\psi \psi}$ this gives an
upper bound on $N_q$. Only for small $q=2,3,4$ or so $N_q$ is larger
than unity, and there is seed forming.

\subsection{Primordial seeds}

The seed $Q$-balls may also be $Q$-balls formed at an earlier
epoch. For this to be possible the initial $Q$-balls should be large
enough to survive the period of evaporation.  The evaporation rate is
given by the detailed balance equation (\ref{balance}).  Ignoring
absorbtion, which is subdominant for $T < T_g$ (note that the
evaporation rate decreases exponentially with temperature), one gets
that the smallest $Q$-ball to survive has charge \cite{frieman}
\begin{equation}
	Q_s \sim 10^{57} \frac{\beta^6}{ g_*^{3/2}} 
	\left( \frac{{\rm GeV}}{m_\psi}
	\right)^{3}  \left(  
	\int^{T_i}_{T_g} \dd T\, \frac{{\rm e}^{-I_Q/T}} {I_Q/T} 
	\right)^3,
\end{equation} 
with $T_i$ the temperature at formation.  For $T_{\rm FO}, T_g < T_i$,
the integral can be approximated by $\sim \exp (-I_Q/T_i)/ (1
+I_Q/T_i)$.  Only for masses $m_\psi \lesssim {\rm eV}$ is $Q_s$
smaller than the total number of particles available in a Hubble
volume at $T_i \sim m_\psi$, eq. (\ref{Q_total}), and is there a
change for very large primordial $Q$-balls to survive the period of
thermal evaporation.

Another possibility is that formation happens at the onset or during
the acretion phase: $T_i \lesssim T_g$.  For large binding energy $I_Q
\to m_\psi$ (which is possible for large $Q$-balls) and large $\eta'$,
accretion dominates over evaporation at non-relativistic temperatures,
see figure~1,

Primordial $Q$-balls may also form during a first order phase
transition~\cite{fggk} from the false ``$Q$-ball vacuum'' to the true
vacuum. At the Ginzburg temperature thermal transitions between
regions of false and true vacuum freeze out; any region of false
vacuum with a charge larger than the minimum charge of a stable
$Q$-ball surviving below this temperature will become a $Q$-ball.  The
potentials under considerations do not exhibit the required first
order phase transition (see footnote 1).  One could add additional
terms to the potential to get a phase transition.  However, the
survival of regions of false vacuum is exponentially surpressed with
size, and correspondingly $Q$-ball formation is exponentially
surpressed with charge.  If formed, the $Q$-balls are expected to be
small $Q \sim Q_{\rm min}$.  Unless there is a mechanism to delay the
phase transition to very low temperatures $T \lesssim 10^{-6} m_\psi$,
these $Q$-balls quickly evaporate and are cosmologically unimportant.

Formation of primordial $Q$-balls through fragmentation of a
condensate~\cite{frag} is studied in the next section.

\section{Bose-Einstein condensation}

We will now study whether there will be condensation.  A condensate
that is unstable under fluctuation can fragment into possibly large
$Q$-balls. We will consider the effective potentials $U_1$ and
$U_2$. In this section $m_\phi=1$, {\it i.e.}, all quantities are
expressed in units of mass.

We will assume that the number density of anti-particles can be
neglected and $\rho \approx n_\phi$.  The state of the system is given
by the minimum of the effective potential for a fixed charge Q:
\begin{equation}
	V(q, \phi) = V(\mu, \phi) + \mu \rho,
\label{Vq}
\end{equation}
with $ V(\mu, \phi)$ the effective potential for a fixed chemical
potential.  In this section $\phi$ denotes the classical background
field, and $\phi_0$ its value at the minimum of $V(q, \phi)$. A
non-zero value of $\phi_0$ signals the existence of a condensate.  At
finite temperature the freqency $\omega$ of the $Q$-ball can be
identified with the chemical potential $\mu$~\cite{laine}.  The charge
density can be solved from
\begin{equation}
	\frac{ {\rm d} V(q, \phi)} {{\rm d} \mu} = 0 \quad 
 	\Rightarrow \quad \rho = \rho(\mu).
\label{rho_mu}
\end{equation}
Eliminating $\mu$ in eq. (\ref{Vq}) then gives the effective potential
in a fixed charge section. A stable configuration lies at the minimum
of $V(q, \phi)$.

To analyze the stability of the condensate one can consider
fluctuations in the homegeneous background.  From the dispersion
relation it follows that fluctuations are amplified for wavelengths
smaller than $k_{\rm max}$~\cite{frag,lee}:
\begin{equation}
	k_{\rm max}^2 = \frac{\rho^2} {\phi_0^4} - U''(\phi_0).
\label{unstable}
\end{equation}
For $\rho^2 - \phi_0^4 U''(\phi_0) < 0$ the above equation does not
have a physical solution and the condensate is stable.

We parametrize the charge density is $\rho = \eta n_\gamma$
\begin{equation}
	\rho \approx 3 \times 10^{-9}\zeta^3 \left( \frac{\Omega_\phi
	h^2}{0.3} \right) \left( \frac{{\rm GeV}}{m_\phi} \right) T^3
	\equiv \eta' T^3.
\label{eta_rho} 
\end{equation}

\subsection{Non-Relativistic Limit}

At zero temperature the charge density $\rho \propto T^3$ is zero, and
there is no condensate.  At non-zero temperature condensate formation
will occur if the charge is larger than the number of excited states.

In the non-relativistic limit the finite-temperature corrections to
the potential are small, and as a first approximation we can use the
zero temperature result $V(\mu,\phi)=U(\phi)-1/2 \mu^2\phi^2$ with
$U(\phi)$ the classical potential, eq. (\ref{U1}, \ref{U2}). Equation
(\ref{rho_mu}) gives $\rho = \mu \phi^2$.  The minimum of $V(q,\phi)$
is at
\begin{eqnarray}
	\rho_1^2 =& \phi_0^4 - 3 A \phi_0^5 + 4 \lambda \phi_0^6,
	&\quad {\rm for} \; U_1
	\nonumber \\
	\rho_2^2 =& \phi_0^4 - 4 A \phi_0^6 + 6 \lambda \phi_0^8,\
	&\quad {\rm for} \; U_2.
\label{rho}
\end{eqnarray}
At low temperatures $\mu \to 1$ and $\rho \approx \phi_0^2 \ll 1$. In
this limit a possible condensate is unstable against decay for values
$\phi_0 < \frac{3A}{8\lambda}$ for $U_1$ and $\phi_0 <
\frac{A}{3\lambda}$ for $U_2$, as follows from eq. (\ref{unstable}).

To see whether a condensate actually forms one has to compute the
density of thermal states. At low temperatures the cubic and quartic
terms in the potential become negligible small, and the theory
approaches the free theory.  In this limit the number of thermal
states is
\begin{equation}
	n_\beta^{\rm NR} = \frac{\zeta(3/2)}
	{(2\pi)^{3/2}} T^{3/2}.
\end{equation}
Since $\rho \propto T^3$, at low temperature all charge will be in the
excited states and the condensate is empty.  The only chance to have a
filled condensate is for $T \to 1$ and $\eta'$ large, so that $\rho >
n_\beta$ or $\eta' T^{3/2} >\zeta(3/2)/(2\pi)^{3/2} \approx
0.17$. Note however that in the limit $T \to 1$ the non-relativistic
approximation breaks down, whereas in the limit $\eta' T^3 \to 1$ the
free field approximation breaks down.

\subsection{Relativistic limit}

We will first consider the potential $U_1$.  The effective potential
for fixed chemical potential to highest order in $T$ is
\begin{eqnarray}
	V(\mu,\phi) = && 1/2 (1 + \lambda T^2 /3 -\mu^2) \phi^2
	- A \phi^3 + \lambda \phi^4 \nonumber \\ && - \mu^2
	\frac{T^2}{6} + c(T) + {\mathcal O}(T),
\end{eqnarray}
with $c(T)$ some temperature dependent constant which we will
drop. From this it follows that
\begin{equation}
	\rho = \mu \phi^2 + \mu T^2.
\end{equation}
The first term in the above equation is the charge in the condensate,
the second term represents the charge in excited states.  The charge
fraction in the condensate is $\phi_0^2/(\phi_0^2 + T^2)$, which is
small for $\phi_0 \ll T$.  The effective potential for fixed charge
density in the relativistic limit becomes
\begin{equation}
	V(q,\phi) =  \frac{1}{2} (1 + \frac{\lambda}{3} T^2) \phi^2
	- A \phi^3 + \lambda \phi^4 
	 + \frac{3 \rho^2}{2(3 \phi^2 + T^2)}.
\end{equation}
Consider the case $\phi_0 \ll T$; then the potential is minimized at
$\phi_0 = 0$ (and thus the approximation is consistent), provided
\begin{equation}
	{\eta'}^2 < \frac{\lambda}{27} + \frac{1}{9 T^2}.
\label{small_eta}
\end{equation}
This can also be seen from the second derivative $V''(q,0) = 1 +
\lambda/3 T^2 - 9 {\eta'}^2 T^2$, which becomes negative for large
$\eta'$.  Thus if condition (\ref{small_eta}) is obeyed there is no
condensate.  For finetuned values of $A^2 / \lambda$ a second minimum
of the potential may develop, but since in the limit of large
temperature the only minimum is at $\phi_0 = 0$ the field will not end
up there.

Consider then the potentially more interesting case that $\eta'$ is
large, and condition (\ref{small_eta}) is not satisfied. Then $V''(0)
<0$ and the potential is minimized at non-zero field value.
Minimization in the limit $T \gg \phi_0$ as well as in the limit $T
\ll \phi_0$ does not yield a consistent solution. It follows that the
minimum is at $\phi_0 \sim T$. This is confirmed by numerical
calculations. The charge density in the condensate is comparable to
that in excited states. The condensate is unstable for $k_{\rm max}^2
> 0$, eq. (\ref{unstable}), with
\begin{equation}
	k_{\rm max}^2 = \frac{{\eta'}^2 T^6}{\phi_0^4}
	 - \frac{1}{3}\lambda(T^2+36 \phi_0^2)
	+ 6 A \phi_0 -1.
\end{equation}
At large temperatures $\phi_0 \propto \lambda^{-1}$ and the second
term in the above equation dominates, as can be verified
numerically. The condensate is stable for large $T$.  The condensate
becomes unstable in the limit $T \to 1$.  As this is also the limit in
which the high temperature expansion breaks down, it is unclear
whether the condensate really fragments.

The analysis for potential $U_2$ is similar.  At high temperature the
effective potential becomes
\begin{eqnarray}
	V(\mu,\phi) = && \frac{1}{2} (1 - \frac{3}{4} A T^2 - \mu^2)
	\phi^2 + (\frac{3}{2} \lambda T^2 - A )\phi^4 \nonumber \\ &&+
	\lambda \phi^6 - \mu^2 \frac{T^2}{6} + c(T) + {\mathcal O}(T).
\end{eqnarray}
For this case, the equivalent of eq. (\ref{small_eta}) is
\begin{equation}
	{\eta'}^2  + \frac{A}{12} < \frac{1}{9T^2}.
\end{equation} 
At large temperature a stable condensate will form with $\phi_0 \sim
\sqrt{A/2+6\eta^2}/2\sqrt{\lambda}$ for small asymmetry or $\phi_0
\sim T$ for large asymmetry $\eta' \gtrsim 1/9$.  The condensate only
survives in the limit $T \to 1$ for large $\eta'$.  The condensate may
be unstable in this limit.

To conclude this section, at non-relativistic temperatures there is no
condensate and all charge resides in excited states.  At temperatures
$T \gtrsim m_\phi$ consensation occurs for large assymetries $\eta'
\gtrsim 1/9$, corresponding to masses $m_\psi \lesssim {\rm eV}$.  The
condensate becomes unstable in the limit $T \to m_\phi$ and fragments
into $Q$-balls. Caution should be taken, as he high temperature
expansion breaks down in this limit.  If the binding energy of the
$Q$-balls is sufficiently large $A^2/2\lambda \gtrsim 10^{-2}$ the
period of evaporation is absent, see figure~1, and these $Q$-balls
survive.

\section{Conclusions}

To summarize, solitosynthesis is a very effecient way to form large
$Q$-balls provided some primordial charge asymmetry and initial seed
$Q$-balls exist.  Most theories do not allow small stable $Q$-ball or
bound state solutions, and solitosynthesis does not start.  The
exception are theories in which the attractive interaction is provided
by a cubic term of the form $A H \psi^* \psi$.  Bound states can form
if the Higgs mass is light ($m_H/m_\psi \lesssim 10^{-3}
{A}^2/m_\psi^2$).  No bound state calculations have been done in the
presence of quartic coupling. We assume that for quartic interactions
that are small compared to interactions governed by the cubic term
bound states persist ($\lambda \lesssim 1$).  We note that if this
assumption turns out to be too optimistic, and stable bound states
require smaller quartic coulings, then small bound states and large
$Q$-balls become mutually exclusive.  This is because for the
potential to admit $Q$-ball solutions the quartic coupling cannot be
too small ($\lambda \gtrsim A^2/m_H^2$).  Succesfull solitosynthesis
will occur if the accretion phase happens before the system falls out
of equilibrium.  All these conditions together limit the paramater
space severely.

For solitosynthesis to have happened in the early universe one needs
$A = 0.1 -1 m_\psi$, at least one of the quartic couplings $\lambda
\sim 1$, $m_H \lesssim 10^{-2} m_\psi$, and ${\rm MeV} \lesssim m_\psi
\lesssim {\rm GeV}$.  This rules out models in which the $H$ field is
the standard model Higgs field, such as the MSSM and the model studied
in \cite{demir}.  The temperature at which $Q$-balls start growing
decreases very rapidly with $A$: $T_g/m_\psi \propto (A/m_\psi)^4$.
For smaller values of the masses or of the cubic coupling than given
above, solitosynthesis may still happen in the future.

$Q$-balls that survive untill present can be part of the the dark
matter in the Universe. For them to play a role during structure
formation they must have been formed before the universe became matter
dominated, that is at temperatures $T_\gamma \gtrsim T_{\rm eq} = 5.5
(\Omega_0 h^2)^{-1} {\rm eV}$. This is only possible for $A'\sim 1$
and $m_\psi \sim {\rm GeV}$.  Whether the $Q$-balls can fullfill the
required cross section to mass ratio to overcome the problems with
cold dark matter as proposed in~\cite{q_dm} remains another question.
More (numerical) studies are needed to determine if solitosynthesis
results in a few $Q$-balls with a very large charge, or in a large
number of $Q$-balls with lesser charge.

Condensate formation is only possible for large asymmetries, or
equivalently $m_\psi \lesssim {\rm eV}$. Symmetries of the order one
can be generated through the Affleck-Dine
Mechanism~\cite{affleck}. Early decoupling increases the number of
charged particles by a factor $\zeta^3$ with $\zeta = (g_{* s}(T_D) -
g_{* s}(T))^{1/3}$, which favors condensation. The condensate becomes
unstable against fluctuations in the limit $T \to m_\psi$, {\it i.e.},
the limit in which all the used approximations break down. Evidently,
better approximations are needed to settle the matter. $Q$-balls
formed through a possible fragmentation of the condensate survive
untill present if accretion dominates over evaporation at
non-relativistic temperatures.  This is possible for $Q$-balls with a
large binding energy, $I_Q = m_\psi - (m_\psi^2 - A^2 / 2 \lambda)^{1/2}
\gtrsim 10^{-2} m_\psi$.

The potentials studied do not allow for a first order phase transition
from the false ``$Q$-ball vacuum'' to the true vacuum.  One could try
and add terms to the potential so that such a phase transition occurs.
However, the $Q$-balls that may form during the phase transition are
small and will evaporate quickly.

Solitosynthesis can lead to a phase transition from the false to true
vacuum.  This will not happen for the potentials studied in this
paper, as for these the field will always end up in the true vacuum.

The author thanks Alexander Kusenko for very helpful discussions.
This work was supported in part by the US Department of Energy grant
DE-FG03-91ER40662, Task C, as well as by a Faculty Grant from UCLA
Council on Research.

%%%%%%%%%%%%%%%%%%%%%%%%%%%%%%%%%%%%%%%%%%%%%%%%%%%%%%%%%%%%%%%%%%%%%%
%% References %%%%%%%%%%%%%%%%%%%%%%%%%%%%%%%%%%%%%%%%%%%%%%%%%%%%%%%%
%%%%%%%%%%%%%%%%%%%%%%%%%%%%%%%%%%%%%%%%%%%%%%%%%%%%%%%%%%%%%%%%%%%%%%

\end{document}